\documentclass[a4paper,10pt]{article}

\usepackage[top=1in,bottom=1in,left=0.75in,right=0.75in]{geometry}
\usepackage{graphicx}
\usepackage{amssymb,amsmath,fancyhdr}

              \title{Synchronized shocks\\ in an inhomogeneous exclusion process}
   \def\shorttitle{Synchronized shocks in an inhomogeneous exclusion process}
           \author{Chikashi Arita} 
\def\shortauthor{C Arita} 
     \def\address{Theoretische Physik, Universit\"at des Saarlandes, 66041 Saarbr\"ucken, Germany}
         \def\abst{We study an exclusion process  with 4 segments, which was recently introduced by T~Banerjee, N~Sarkar and A~Basu [J. Stat. Mech. (2015) P01024]. The segments have hopping rates 1, $r(<1)$, 1 and $r$, respectively. In a certain parameter region, two shocks appear, which are not static but synchronized.  We explore dynamical properties of each shock and correlation of shocks, by means of the so-called second-class particle. The mean-squared displacement of shocks has three diffusive regimes, and the asymptotic diffusion coefficient is different from the known formula. In some time interval, it also  exhibits  sub-diffusion, being proportional to $ t^{1/2} $.  Furthermore we introduce a correlation function and a crossover time, in order to quantitatively characterize the synchronization. We numerically estimate the dynamical exponent for the crossover time.  We also revisit the 2-segment case and the open boundary condition for comparison.}
              \date{   }

\makeatletter
\def\@maketitle{ 
\begin{center} 
  \let \footnote \thanks
    {\LARGE\linespread{1.2}\selectfont\textbf{\@title}\par}  \vskip 10mm 
    {\Large \@author}           \vskip 5mm  
    {\address}    \vskip 5mm 
    {\large \@date}              \vskip 5mm 
    \textbf{Abstract}     \end{center}
  \begin{quote} \abst \end{quote}   \vskip 5mm  
\noindent\makebox[\linewidth]{\rule{\textwidth}{0.5pt}}}
\makeatother

\fancypagestyle{titlepage}{
 \lhead{}  \chead{}  \rhead{}
 \lfoot{}  \cfoot{\thepage}  \rfoot{}    }

\begin{document}

\maketitle 

\thispagestyle{titlepage}

\section{Introduction}\label{sect:intro} 
The exclusion process is a paradigm for non-equilibrium behaviour \cite{bib:M1}. 
In the one-dimensional totally asymmetric simple exclusion process (TASEP),
each site $i$ $ ( i=1,2,\dots, L) $ is either occupied by one particle ($ \tau_i =1 $) or empty ($ \tau_i =0$),
and a particle at site $ i$ hops to $i+1 $ with rate 1, if the target site is empty. Investigation of shocks is one of central issues in the TASEP\cite{bib:BCFG,bib:FF,bib:DJLS,bib:DEM,bib:M2,bib:KSKS,bib:CHA}. Let us recall a known result in the TASEP with open boundaries (shortly, open TASEP), where a particle is injected at site $i=1$ with rate $\alpha$, and extracted at site $i=L$ with rate $ \beta $ \cite{bib:DEHP}. These rates are regarded as reservoir densities $\alpha$ and $ 1-\beta $, respectively. The case $ \alpha = \beta < 1/2 $ is called the co-existence line. Two plateaus with densities $ \alpha$ and $ 1-\alpha$ co-exist in the domains
 $ 1 < x < S(t) $ and $ S(t) < x < L $, respectively. The position $ S (t) $ of the shock (domain wall) is \textit{dynamical}, and its  behaviour is diffusive \cite{bib:DEM,bib:KSKS}, \textit{i.e.}  
\begin{align} \label{eq:D=} 
    \big\langle \big( S (t) - S (0) \big)^2 \big\rangle_{\mathrm E} \simeq   2 D (\alpha ) t,   \    
     D (\alpha ) = \frac{ \alpha(1-\alpha) }{ 1 - 2\alpha } , 
\end{align} 
as $t\to \infty$. Here $ \big\langle \cdot \big\rangle_{\mathrm E} $ denotes the ensemble average. 
This asymptotic diffusion coefficient is also true on $ \mathbb Z $ \cite{bib:FF}.

\begin{figure}\begin{center}
     \includegraphics[width=0.24\textwidth]{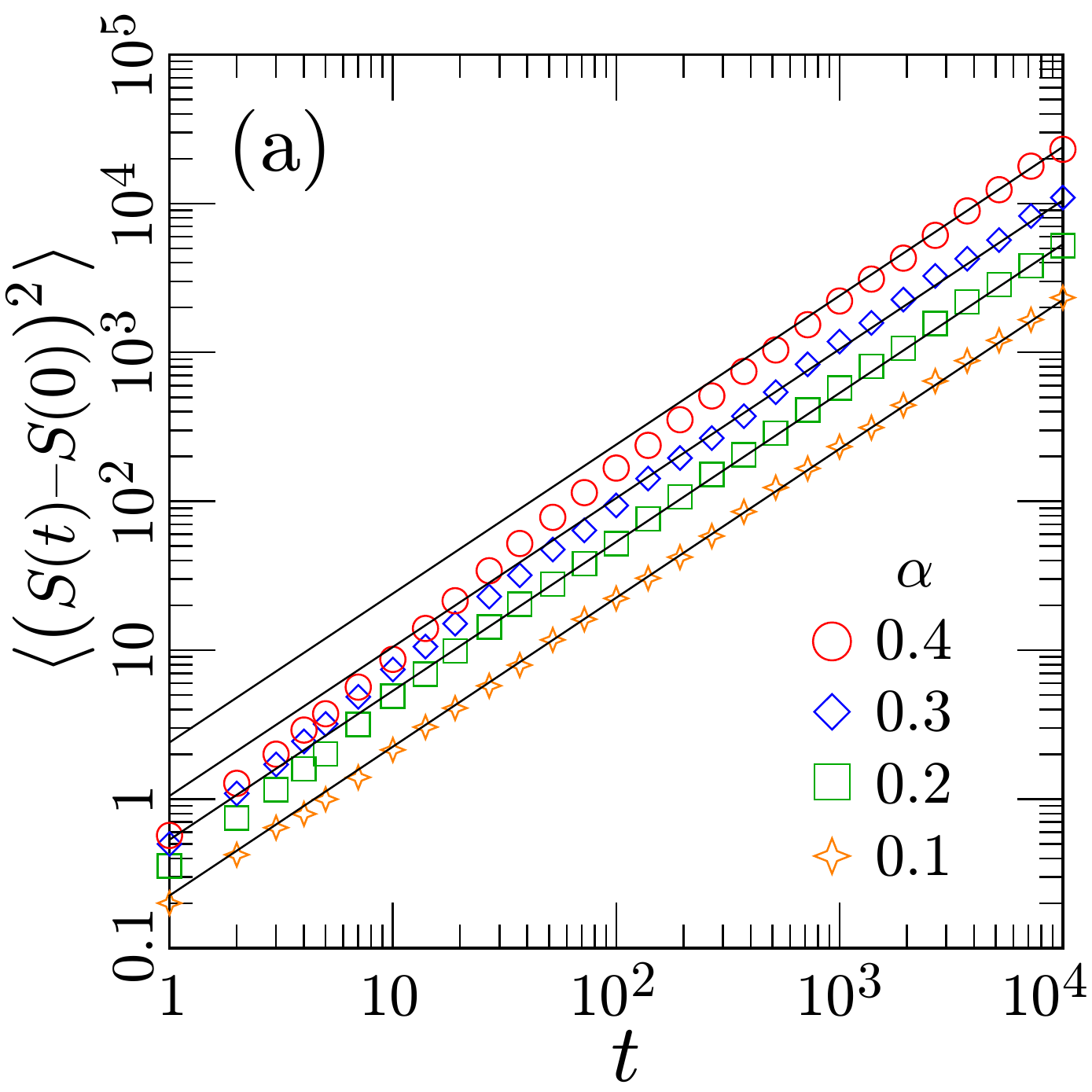} \  
     \includegraphics[width=0.24\textwidth]{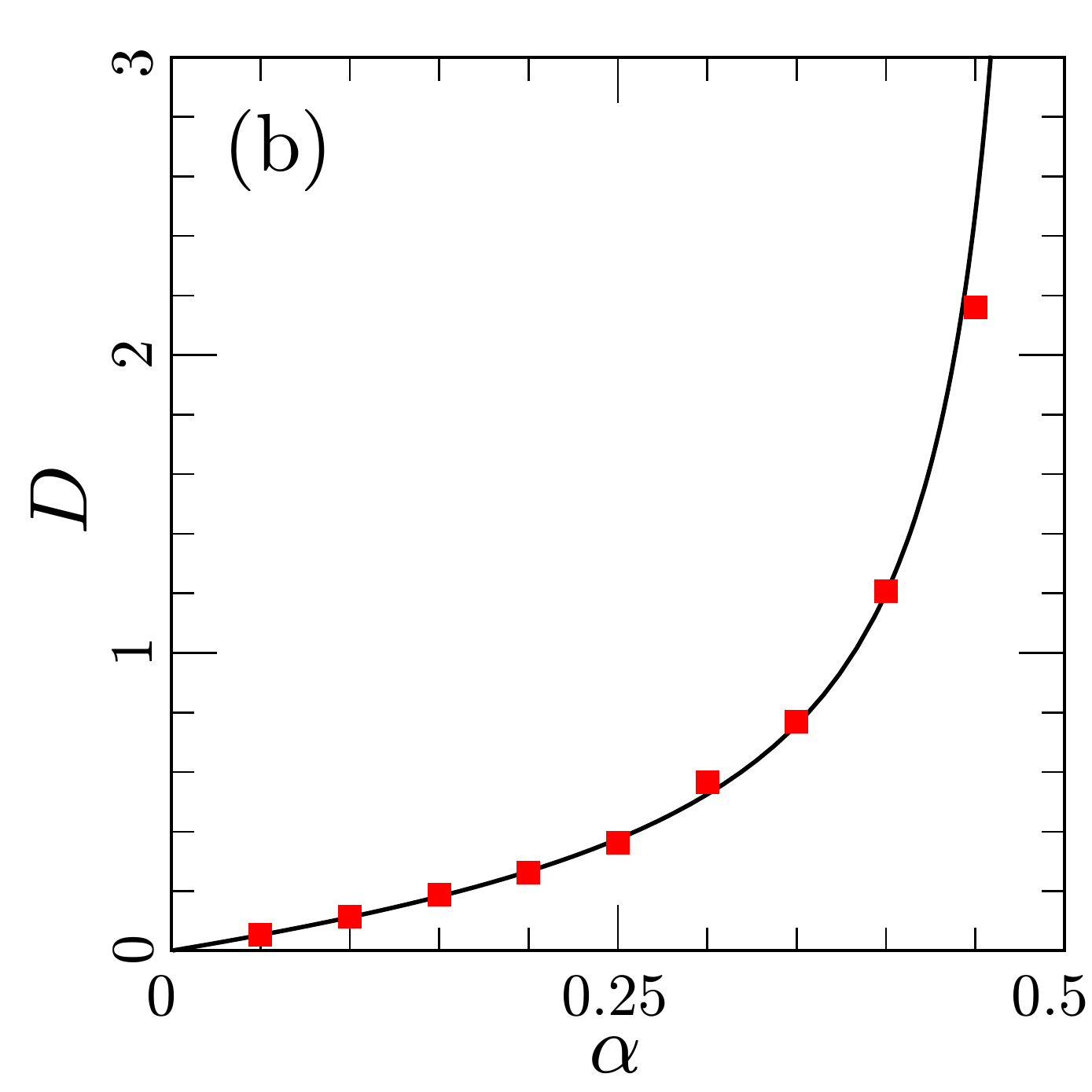} \end{center}
\caption{ 
  (a) MSD of the second-class particle vs time, and
  (b) diffusion coefficient vs boundary rate 
      on the co-existence line of the open TASEP with $ L=10^4 $. 
 The markers in (a) are simulation results averaged over  $ 10^3 $ runs.
  We used simulation data  $  ( S(t) -S(0) )^2  / (2t ) $ in $  5\times 10^3 \le  t \le 10^4  $ to plot the markers in (b).   The lines in (a,b) correspond to eq.~\eqref{eq:D=}.   
 \label{fig:openTASEP}} 
\end{figure}

The so-called second-class particle ($ \tau_i =2 $)  microscopically \textit{defines}  
 the positions of shocks \cite{bib:BCFG}. It behaves as a hole for particles, and as a particle for holes, \textit{i.e.}  the  hops $ 20  \to  02 $ and  $ 12 \to  21 $ occur between sites $i$ and $ i+1 $, with the same rate   as for $ 10 \to 01 $. We confine only one second-class particle into the system in the initial state. It is not extracted from the system and we do not inject another second-class particle, \textit{i.e.}  
``semi-permeable boundary condition'' \cite{bib:A1,bib:A2,bib:U,bib:ALS}. Figure \ref{fig:openTASEP} (a,b) shows   simulation results of the mean-squared displacements (MSD) of the second-class particle, which agree with the formula  \eqref{eq:D=}. We notice that finite-time effect becomes strong, when $\alpha$ approaches $1/2$. The density profile averaged over a large time interval 
\begin{align}
  \rho_i := \langle \tau_i  ( 2-\tau_i ) \rangle_{\mathrm T}    
  \label{eq:rho_i:=}
\end{align}
looks different from snapshots in simulations. 
Since the shock position moves evenly in the chain,  $ \rho_i $ is given by 
\begin{align} \rho_i \simeq ( 1 - 2\alpha ) \frac i L  + \alpha, \label{eq:rho_i=open} \end{align}
which was shown by the exact stationary state  \cite{bib:DEHP}.

On the other hand,  static shocks are realized in the TASEP, by imposing  attachment and detachment of particles in the bulk of the chain (the Langmuir kinetics \cite{bib:PFF}).   Static shocks  were also found in  TASEPs with inhomogeneous hopping rates on a ring.     One of the simplest cases    is  the Janowsky-Lebowitz (JL) model \cite{bib:JL}: 
 \begin{align}    p_i =     1  \  ( 1 \le i < L )   , \  r<1 \  ( i=L )  ,  \end{align} 
 where  $ p_i $  denotes the hopping  rate from site $i$ to $i+1$ ($ L+1 := 1 $). 
 Due the inhomogeneity of the bond between sites $ L $ and 1,  the JL model exhibits a shock profile in a certain parameter region.  A mean-field theory qualitatively explains a phase transition between shock and flat density profiles, which  is still a fascinating problem \cite{bib:CLST,bib:SPS}. 
 
In \cite{bib:TB},  another inhomogeneous TASEP   was introduced: 
\begin{align}\label{eq:two_pi=}  p_i =   1  \  ( 1 \le i \le \ell    )   , \  r  \  ( \ell <  i \le 2 \ell=L   )  .\end{align}
 We refer to this model as 2-segment TASEP.   In a certain region of the parameter space $ (r,\rho) $ (where $ \rho$ is the global density \textit{i.e.} the number of particles $/L$), this model also exhibits a static shock.  Recently  the authors of  \cite{bib:BSB} investigated TASEPs with three and four parts, generalizing \eqref{eq:two_pi=}.  Here, we mainly study a specific  4-segment TASEP 
\begin{align}\label{eq:four_pi=}
  p_i = 
  \begin{cases}  
     1 & ( 1 \le i \le \ell   \ \vee \   2\ell  < i \le  3\ell  )   , \\ 
     r & (  \ell < i \le 2\ell \  \vee\   3\ell < i \le 4\ell =L) .    \\ 
  \end{cases}
\end{align}
There can exist two shocks in  the 1st and 3rd segments. The positions of the shocks cannot be fixed even in the macroscopic level, but they are related to each other by an equation derived by the particle number conservation  \cite{bib:BSB}. 
 The purpose of this work is performing more detailed Monte Carlo simulations (in continuous time),  in order to deeply understand this synchronization phenomenon. 

Before investigating the 4-segment TASEP, we reconsider the case of 2 segments via the second-class particle.  As an evidence that the second-class particle indicates the shock position, we check that its spatial distribution becomes gaussian, corresponding to the density profile written in the error function.  We also examine properties of the standard deviation of the shock position. 
Then we turn to  the 4-segment TASEP. We find that the MSD exhibits various behaviours, depending on the time scale that we focus. The open boundary condition that we have already reviewed is the reference case. We quantify the interaction between the  shocks by a correlation function.
 Furthermore we introduce a crossover time distinguishing between time scales  of   independency and synchronization of shocks. We also  numerically  estimate the dynamical exponent of the crossover time. Finally we give the conclusions of this work and some remarks, including possible future studies.  Overall in this work, we use the following  definition  for  macroscopic density profiles  with  \textit{mesoscopic} length  of the lattice  $ 2 \delta +1 $ ($ 1 \ll \delta \ll  \ell $),  which is in general different from  the microscopic density profile \eqref{eq:rho_i:=}:
\begin{align}
   \rho (x)  = \rho ( i/\ell   )  = 
    \frac{1}{ 2 \delta + 1 } \sum_{ i' = i - \delta }^{ i + \delta }  \tau_{i'}  ( 2 - \tau_{i'}  )  .   \label{eq:macro-density}
\end{align}

\section{2-segment TASEP}\label{sect:2-seg}
 Let us consider the model \eqref{eq:two_pi=}.  
 We begin with the assumption that the global density $ \rho $ is enough small, and 
 each segment has a flat density profile       
\begin{align}\label{eq:rho(x)=12}
   \rho(x) =      \alpha_1   (  0  < x < 1   )   , \       \alpha_2   (  1 < x < 2  )   .    
\end{align}
The conservation laws of the number of  particles  and the stationary current $ J $ are written as 
\begin{align}
  \label{eq:2rho=a1+a2}
  2 \rho &=     \alpha_1   +     \alpha_2  ,  \\  
        J &=  \alpha_1 (1-\alpha_1) = r  \alpha_2 (1-\alpha_2)  , 
  \label{eq:J=J1=J2}
\end{align}
respectively. 
These are easy to solve \cite{bib:TB}: 
\begin{align}
\label{eq:alpha1=,alpha2=}
  \alpha_1  &= \frac{ 1 + r -4r\rho - R }{  2 (1-r)  } , \ 
  \alpha_2  = \frac{ 4\rho - (1+r)  + R}{  2 (1-r)  } , \\
  J &=          \frac{ r(1-2\rho)(R-(1+r)(1-2\rho)  ) }{   (1-r)^2  } 
    \label{eq:J:LDLD}  
\end{align}  
 with $ R  = \sqrt{(1+r)^2-16r\rho (1-\rho)}  $. 
The density profile \eqref{eq:rho(x)=12} is realized as long as  $ \alpha_2 <  1/2  $
with \eqref{eq:alpha1=,alpha2=}, \textit{i.e.}  
\begin{align}  \label{eq:rho_c=} \rho <   \rho_c : =   \frac { 2-\sqrt{1- r} }{4}     .
   \end{align}
Since both $ \alpha_1 $ and $  \alpha_2 $ are less than $ 1/2 $, 
 the case $ \rho < \rho_c $ is called LD-LD phase (LD$=$low density).

\begin{figure}\begin{center}
  \includegraphics[width=0.24\textwidth]{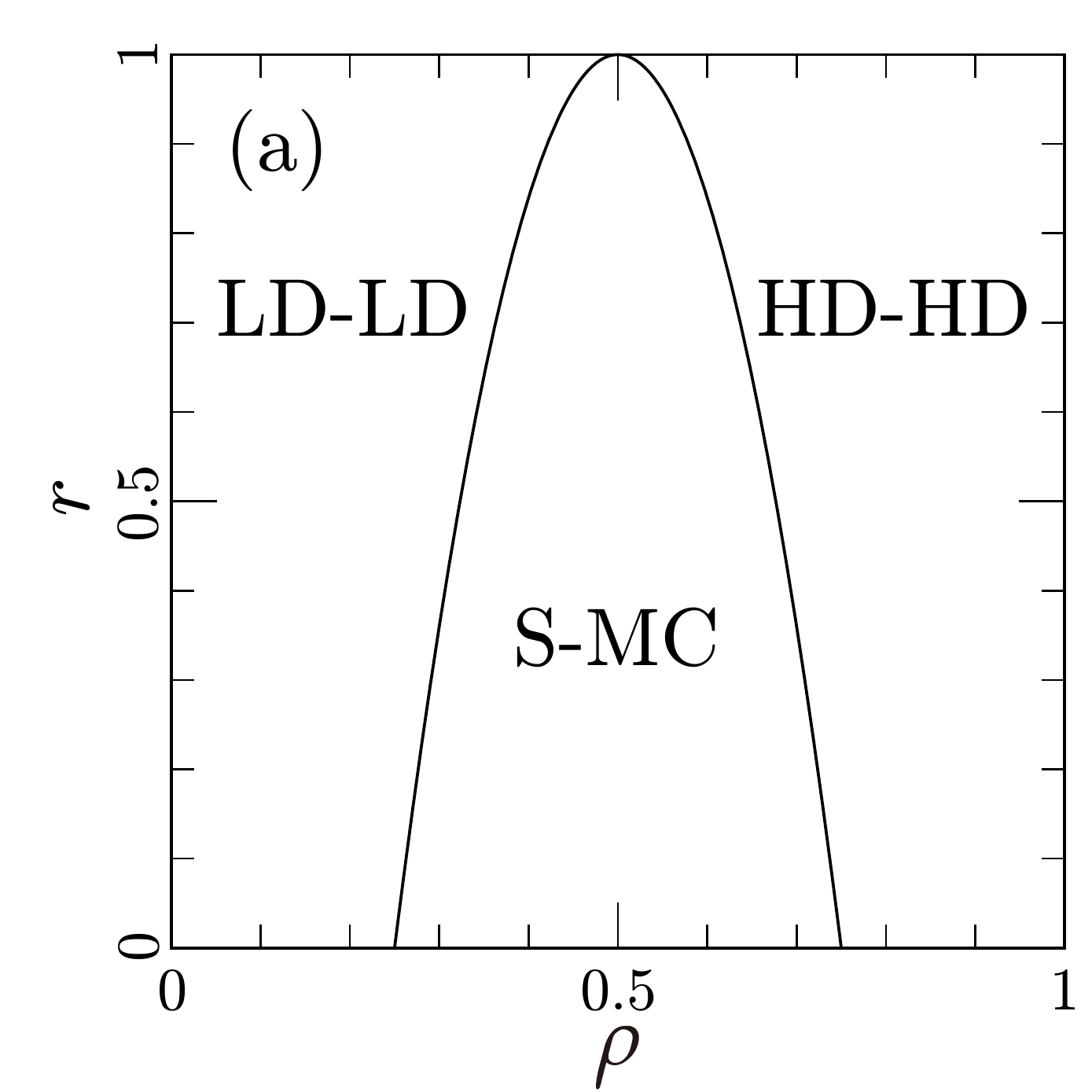} \ 
  \includegraphics[width=0.24\textwidth]{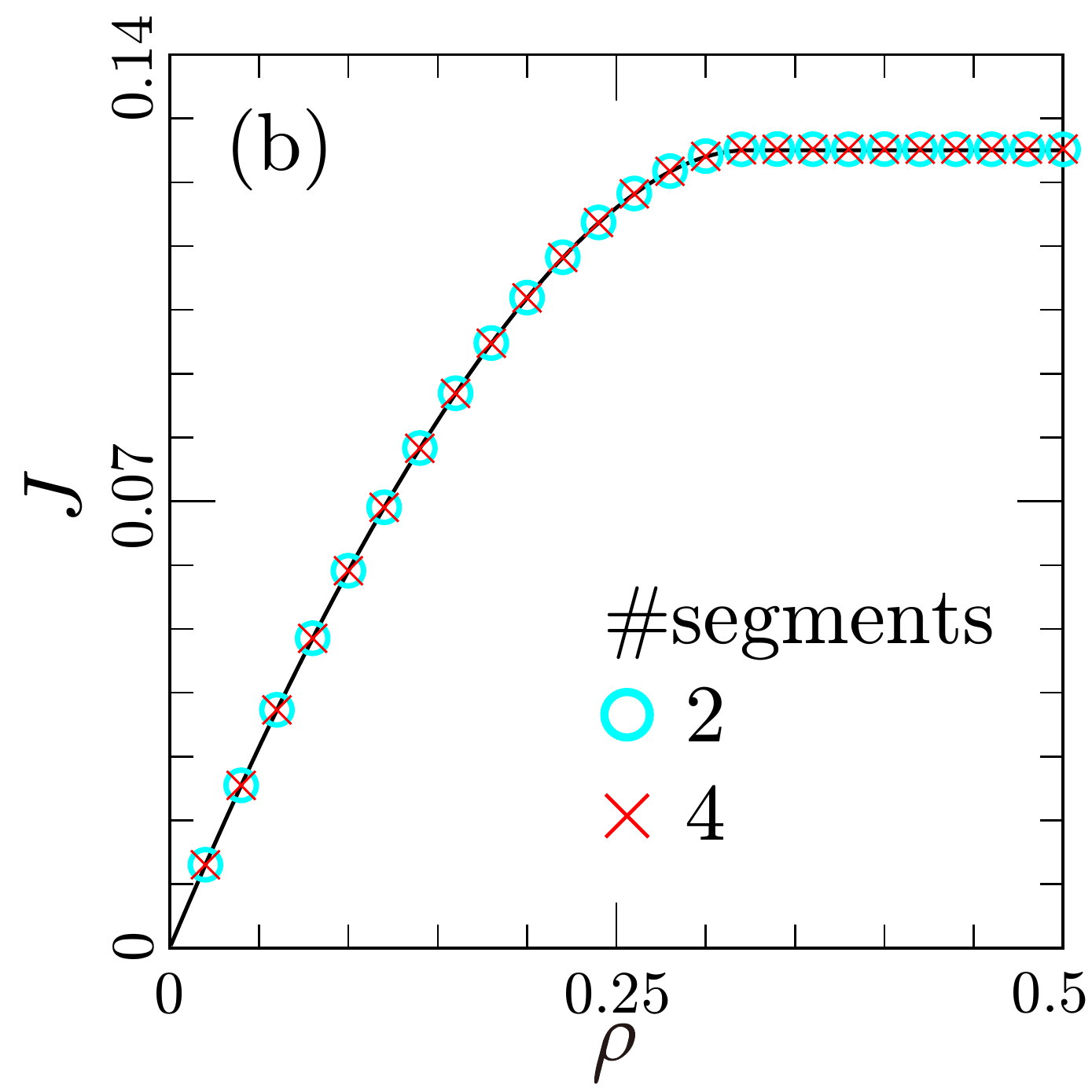} \end{center} 
\caption{  
 (a) Phase diagram  and
 (b) fundamental diagram 
    of the 2- and 4-segment TASEPs \cite{bib:TB, bib:BSB}. 
The phase boundaries in (a) are given by $ \rho = \rho_c ,1 - \rho_c$  
\textit{i.e.}  the parabola  $ r =  1  -  1 6  ( \rho - 1/2   )^2    $.  
For  (b),  we have set    $ \ell = 10^3 $ and $ r =1/2$.  To obtain each marker, we counted the number of flowing particles in a single simulation run, and took average over $ 10^6 \le t \le 10^7 $. The line in (b)  corresponds to the   predictions  \eqref{eq:J:LDLD}, \eqref{eq:J:SMC}. 
  \label{fig:diagrams}} 
\end{figure}

When  the total density $  \rho $ exceeds the critical density, 
the 2nd segment maintains  the density $1/2 $ and
a shock appears on the 1st segment: 
\begin{align}\label{eq:rho(x)=1lh2}
   \rho(x) = 
  \begin{cases}  
     \alpha_1  \  (  0  < x < s )   , \  
     1- \alpha_1 \  (  s < x < 1   )   , \\
     \alpha_2 =   \frac{1}{2}   \  (  1 < x < 2  )  .    
  \end{cases}
\end{align}
We refer to this case as S-MC (shock-maximal current) phase. The shock position $s$ is macroscopically static, \textit{i.e.}  localized.  By solving the conservation laws of the number of particles and the stationary current 
\begin{align}
 2 \rho =&  s \alpha_1  + (1-s)(1 - \alpha_1 ) +  \  \alpha_2  ,  \\
 J  ( \alpha_1 ) =& \alpha_1 (1-\alpha_1) =  r \alpha_2 (1-\alpha_2 )  =   r / 4   \label{eq:J:SMC} , 
\end{align}
the densities and the shock position in the 1st segment  are determined as 
\begin{align}
  \alpha_1 = \frac{1-\sqrt{1-r}}{2}  ,  \    s  = \frac{1}{2} + \frac{1-2\rho}{ \sqrt{1-r}}  . 
  \label{eq:alpha1=,s=}
\end{align}
The parameters $ ( r , \rho ) $ are inversely specified by $ (\alpha_1,s) $ in the S-MC phase. We have emphasized the current as a function of the density $ \alpha_1 $. 
 
When the global density $\rho$ exceeds  $1-\rho_c $, the shock position reaches site $i=1 $. The densities in both segments are flat, and larger than $ 1/2 $, \textit{i.e.}  the HD-HD phase (HD$=$high density). Figure \ref{fig:diagrams} (a) summarizes the three phases. In fig.~\ref{fig:diagrams} (b), we check that the predicted currents \eqref{eq:J:LDLD} and \eqref{eq:J:SMC} are realized by simulations. Because of the particle-hole symmetry, we restrict our consideration to $ \rho \le 1/2 $.

\begin{figure}\begin{center}
     \includegraphics[width=0.24\textwidth]{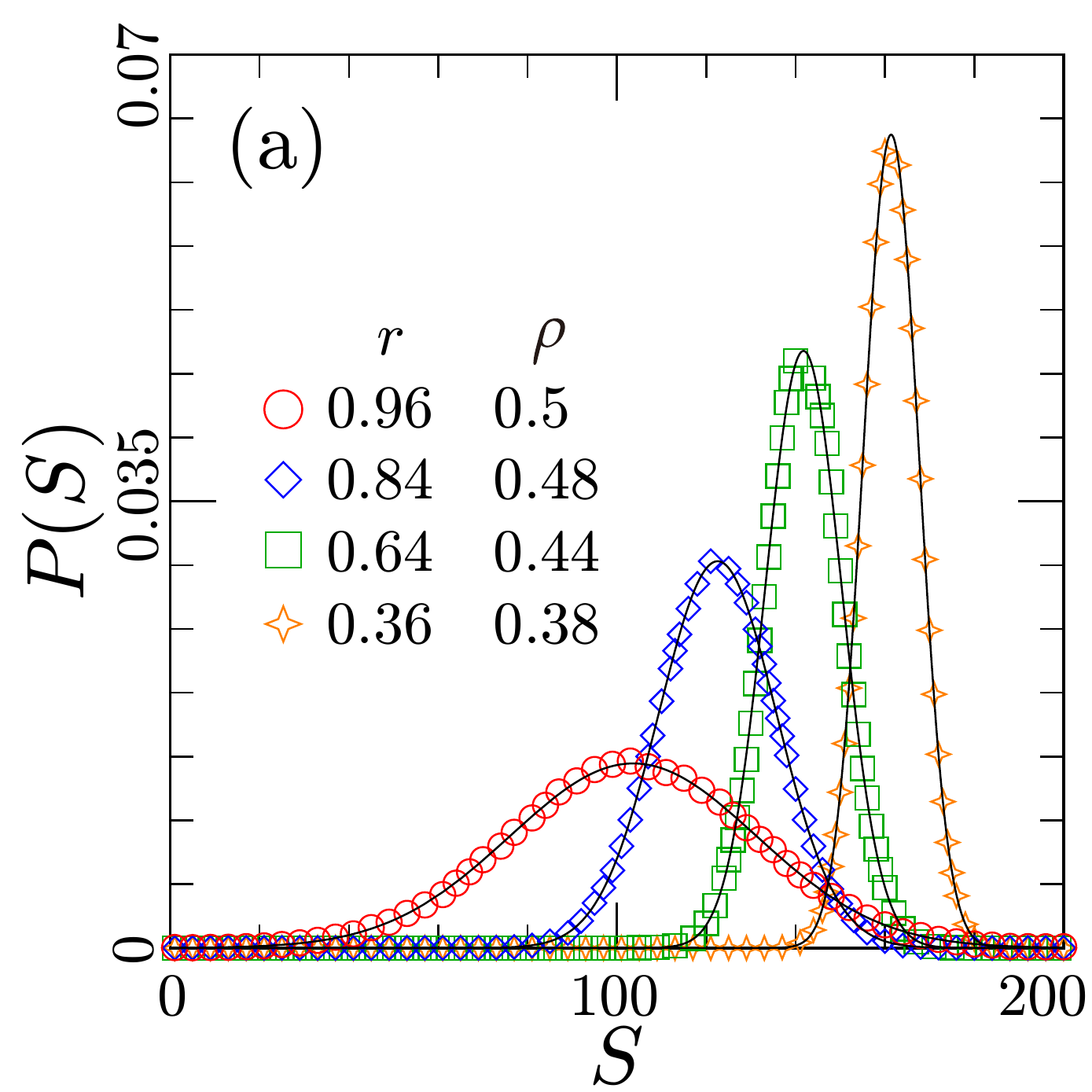}  \ 
     \includegraphics[width=0.24\textwidth]{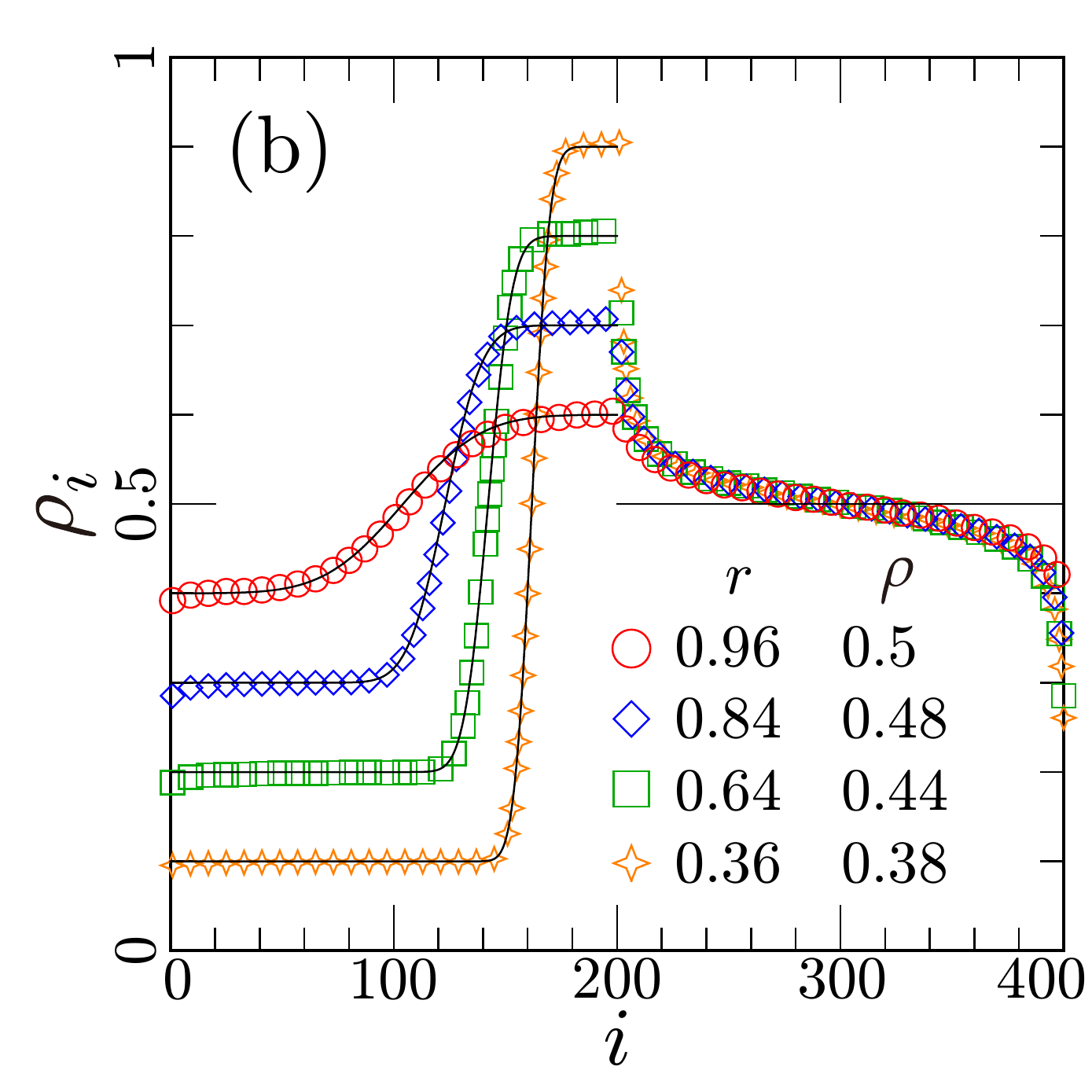} \  
     \includegraphics[width=0.24\textwidth]{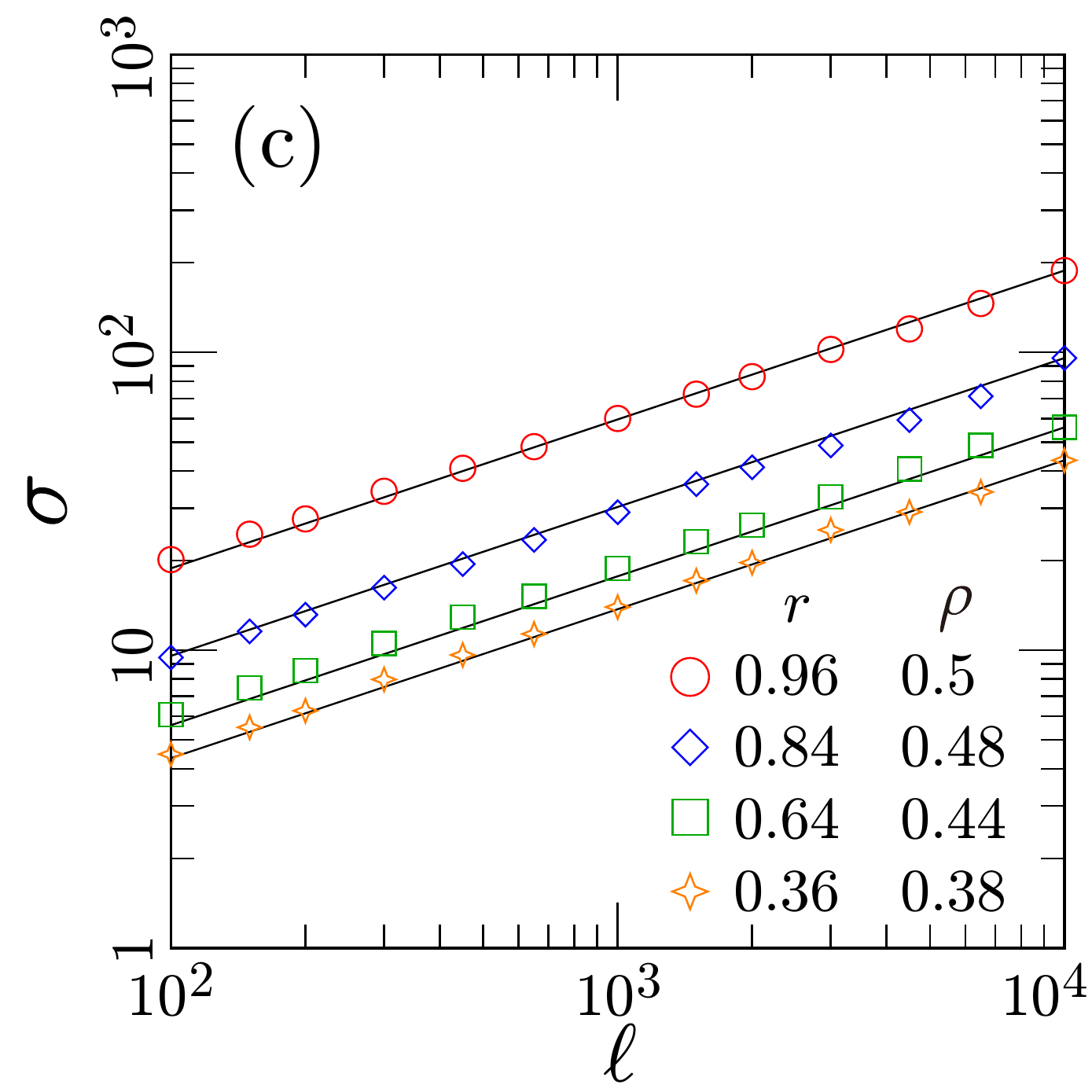} \ 
     \includegraphics[width=0.24\textwidth]{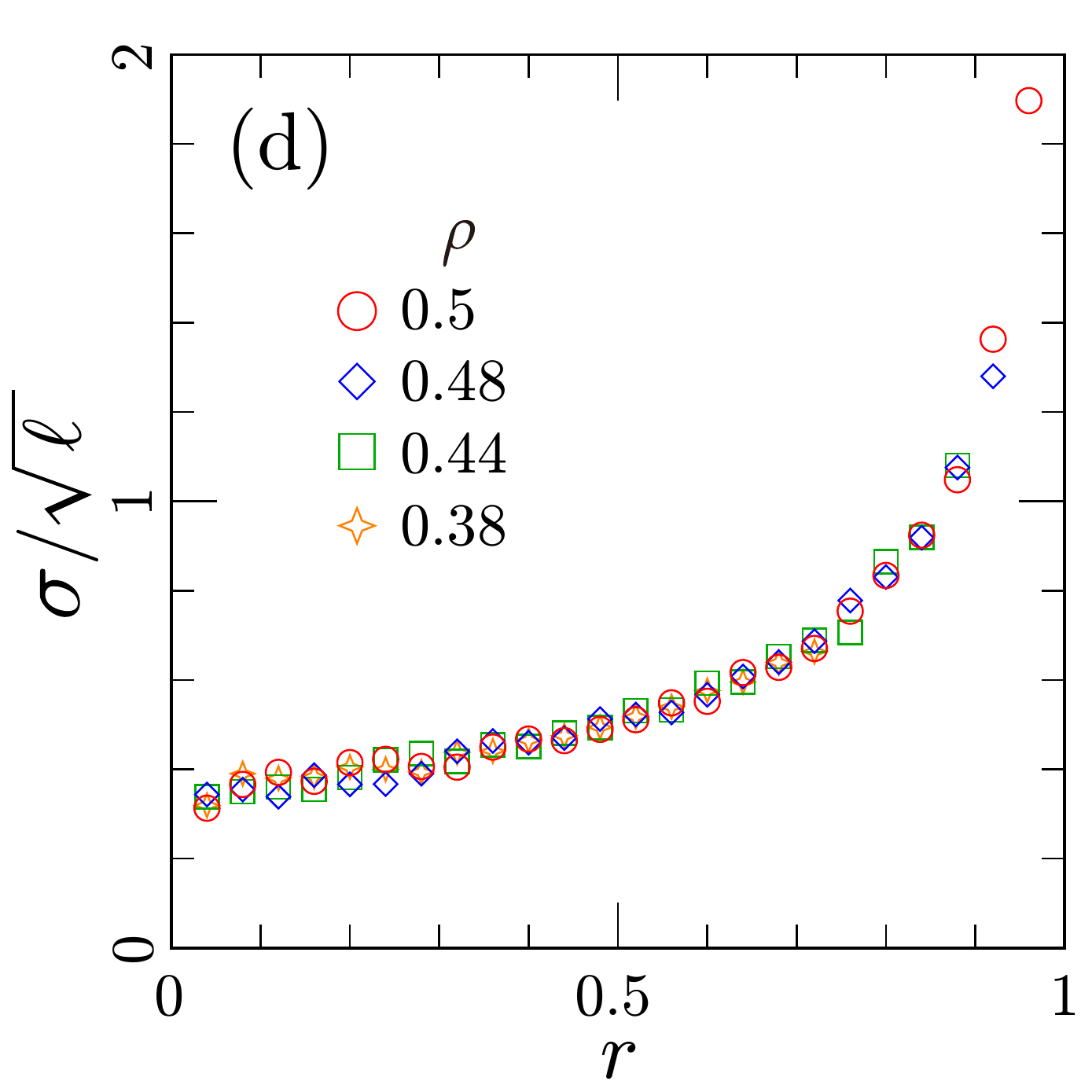} 
     \end{center}  
\caption{  
 (a) Distributions of the second-class particle, 
 (b) density profiles,
 (c) standard deviation $ \sigma $  vs segment length   $ \ell $,  and  
 (d) $\sigma$ vs hopping rate $r$.   
 For (a) and (b) we  have set  the system length as $ \ell =200$, and for (d) $ \ell =10^3 $.   
  Each  plot marker was obtained by averaging  data of a single simulation run  over 
   $ 10^6  \le t \le 10^7$ 
  [exceptionally $ 10^6  \le t \le 5\times 10^7$ for the cases $ \ell > 10^3 $ in  (c)]. 
   The curves  in  (a) and (b) correspond to  
    eqs.~\eqref{eq:P(S)=Gauss}  and  \eqref{eq:<tau>=erf},  
    respectively, with $ \langle S \rangle  $   and $ \sigma$ obtained from simulation data. Each line  in (c)  corresponds to  $ ( \ell  /10^4 )^{1/2}    \times  \sigma |_{\ell=10^4} $ with $ \sigma |_{\ell=10^4} $ obtained by simulations. 
\label{fig:2-seg} } 
\end{figure}

Let us investigate properties of the shock in the S-MC phase in more detail. There is a microscopic deviation of the density profile near the shock position. It obeys a gaussian distribution, \textit{i.e.}  the probability distribution $ P ( S ) := \frac 1 2 \langle \tau_S (\tau_S-1) \rangle_{\mathrm T} $ of the position $S$ of the second-class particle is given as 
\begin{align}
\label{eq:P(S)=Gauss}
 P (S)  =   \frac{1}{ \sqrt{2 \pi  \sigma^2  }   }  
  \exp  \bigg[  - \frac{1}{ 2\sigma^2 }  ( S - \langle S \rangle_{\mathrm T}  ) ^2  \bigg]     
\end{align}
with $ \langle S \rangle_{\mathrm T}  \simeq  \ell s $ $ ( \ell \to \infty ) $. We see good agreement of simulations to this distribution in fig.~\ref{fig:2-seg} (a).  This corresponds to the fact that the (microscopic) density profile \eqref{eq:rho_i:=} in the 1st segment  is well described in terms of  the error function  $\mathrm{erf}[\cdot]$  \cite{bib:BSB}, see  fig.~\ref{fig:2-seg} (b): 
\begin{align}
  \label{eq:<tau>=erf}
    \rho_i =  \frac{\sqrt{ 1-r } }{2}  \ \mathrm{erf} 
   \bigg[  \frac{ i -  ( \langle S \rangle_{\mathrm T}  +  \frac 1  2  )  }{ \sqrt{2   \sigma^2  } } \bigg] 
       +     \frac{1}{2}   .
\end{align} 
We chose the values of  parameters, such that corresponding  densities and shock positions are given as  
\begin{align}
  \nonumber 
 &    ( r , \rho   ) = (0.96,  0.5  ) , ( 0.84 , 0.48 ) , ( 0.64,0.44 ) , (0.36,0.38  )   \Leftrightarrow  \\  
 &    ( \alpha_1 , s  ) =  ( 0.4 , 0.5 ) , (0.3, 0.6   ) ,  ( 0.2 , 0.7  ) ,  ( 0.1, 0.8  ) , 
\end{align}
respectively,  according to eq.~\eqref{eq:alpha1=,s=}. 
(In the 2nd segment, we expect that the density profile, which is deviated from $  1/2 $,  can be well written by the exact finite-size effect in the maximal current phase of the open TASEP \cite{bib:DEHP}.)

 In fig.~\ref{fig:2-seg}  (c), we observe $ \sigma \propto \sqrt \ell $  \cite{bib:BSB}.  Furthermore  $ \sigma / \sqrt \ell   $ vs $r $ is shown in fig.~\ref{fig:2-seg}  (d) with different global densities  $ \rho $.  So far we have not  found an explicit formula of $ \sigma / \sqrt \ell $, but it seems independent of $ \rho $.

\section{4-segment TASEP}\label{sect:4-seg}
\begin{figure}\begin{center} 
    \includegraphics[width=0.24\textwidth]{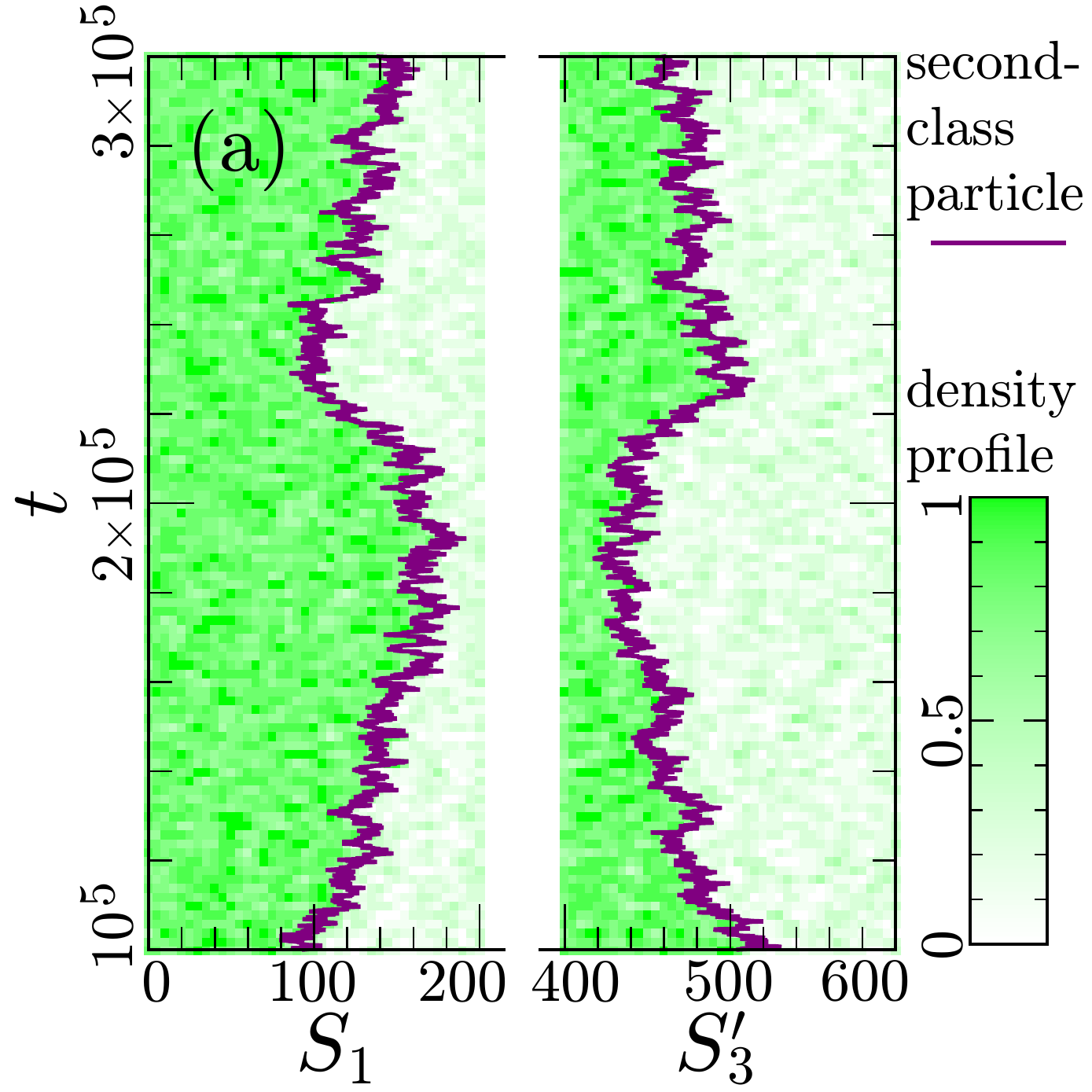} \ 
    \includegraphics[width=0.24\textwidth]{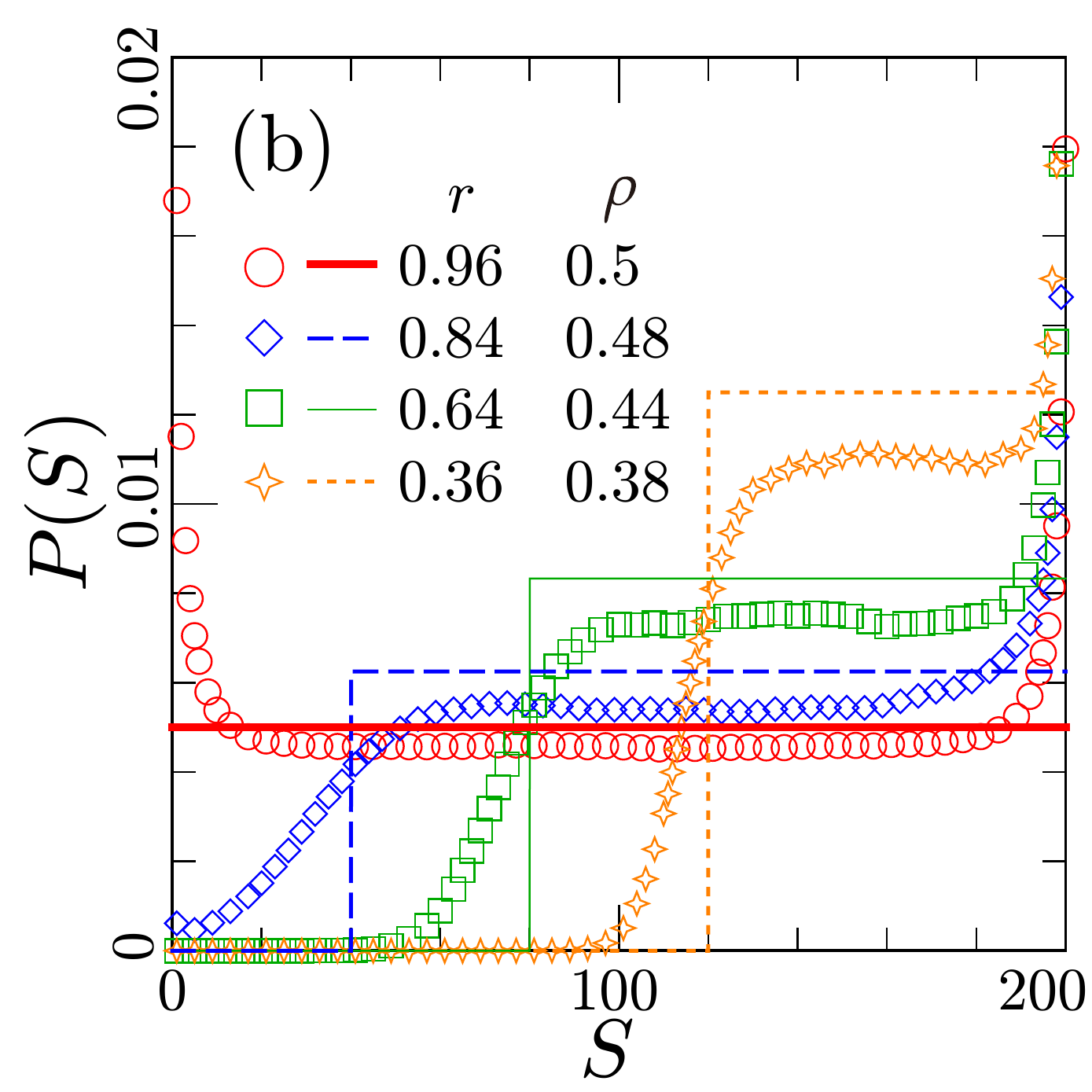} 
    \includegraphics[width=0.48\textwidth]{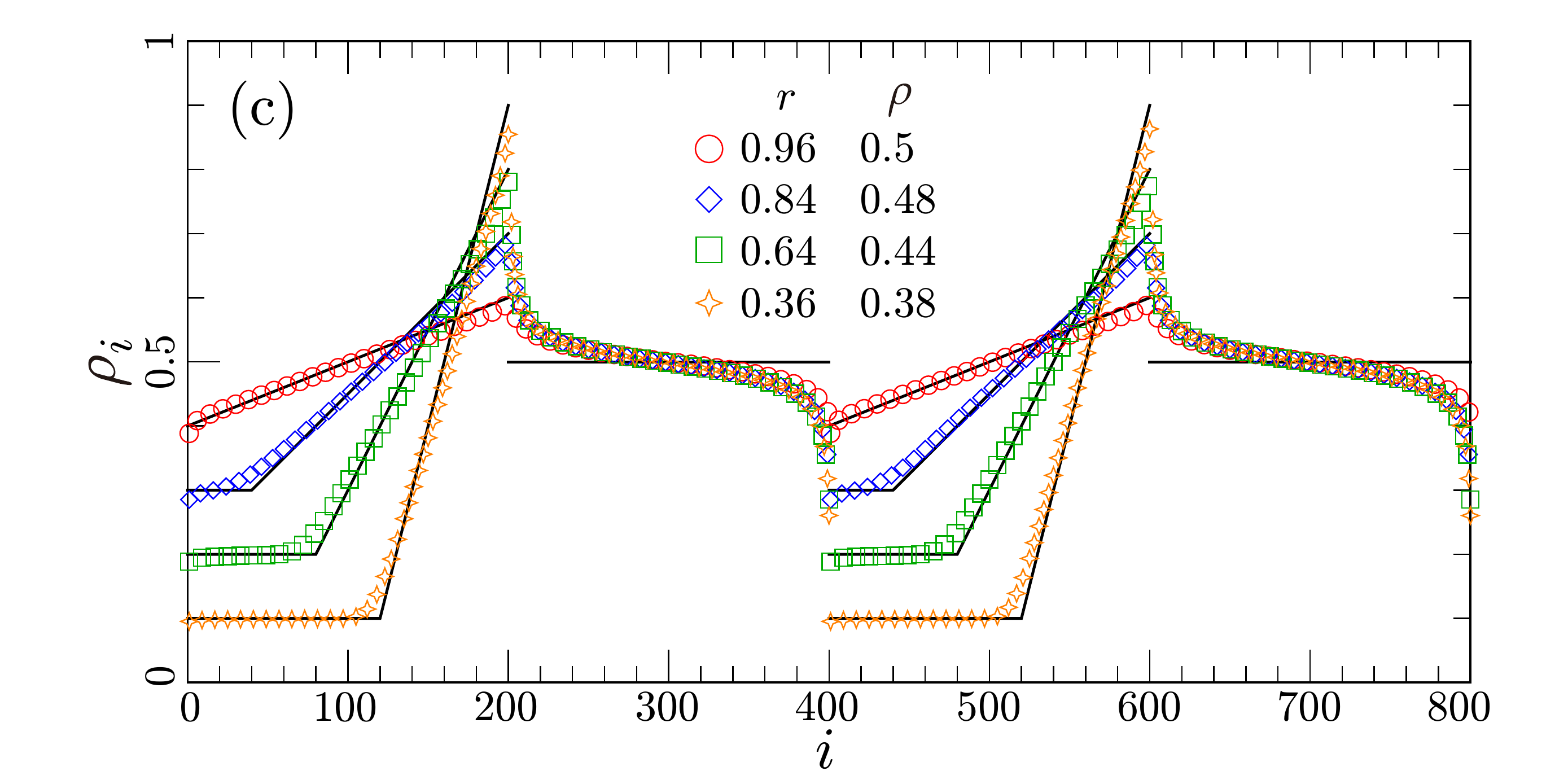} 
    \end{center} 
\caption{  
 (a)  Kymograph probing  
   second-class particles and macroscopic density profiles
    [eq.~\eqref{eq:macro-density}   with   $\delta=5$]  
    in the 1st and 3rd segments 
    for $ (r,\rho ,\ell ) = (0.64, 0.5,200 )  $.   
 (b) Distribution  of the second-class particles, and (c) density profiles
     for $ \ell =200$, averaged over      $ 10^6 \le  t \le 10^8 $.
 The lines in (b)  are the uniform distribution over the range \eqref{eq:lambda<sj<1}, and in (c) the prediction \eqref{eq:rho(x)_SMCSMC_over-time} with $ x = i /\ell    $. 
\label{fig:4-seg} } 
\end{figure}

Now we turn to the 4-segment TASEP \eqref{eq:four_pi=} \cite{bib:BSB}. By the same argument  as for the 2-segment case,  the density $\alpha_j  $ of each segment $j$ is flat, when  $ \rho < \rho_c  $. (The phase transition line $ \rho = \rho_c $ is identical to that of the 2-segment case, fig.~\ref{fig:diagrams} (a).)
 From the translational invariance,  we have  $ \alpha_1 = \alpha_3  $ and  $ \alpha_2 = \alpha_4  $.  Furthermore these densities have  the same form  \eqref{eq:alpha1=,alpha2=}  as for the 2-segment case \cite{bib:BSB}. 
 After the global density exceeds $ \rho_c  $, the 2nd and 4th segments maintain the density $\alpha_2 =\alpha_4 = 1/2 $, and shocks appear in the 1st and 3rd segments \cite{bib:BSB}:
  with   $ \alpha_1 = \alpha_3   $,  
\begin{align}\label{eq:rho(x)_SMCSMC}
   \rho(x) =   
   \begin{cases}  
     \alpha_1  \ (  0  < x < s_1 )   ,    \    1- \alpha_1 \  (  s_1 < x < 1   )   , \\ 
        \frac{1}{2} \   (  1 < x < 2  )  ,  \\ 
     \alpha_3  \ (  2  < x < s_3'   )   ,    \    1- \alpha_3 \  (  s_3'  < x < 3    )   , \\ 
        \frac{1}{2} \   (  3 < x < 4   )    . 
  \end{cases}
\end{align}
The shock positions are denoted by  $ s_1 $ and $ s_3'   = s_3 +2$  [$ s_3=0 $ (resp. $ s_3=1  $) corresponds to the boundary between 2nd and 3rd  (resp.  3rd and 4th) segments]. 
 The conservation of the number of particles is written as 
\begin{align}
 4  \rho = 
  s_1 \alpha_1  + (1-s_1) (1-\alpha_1 )  +   \alpha_2 
   +   s_3 \alpha_3  + (1-s_3)(1- \alpha_3 ) +   \alpha_4   .
\end{align}
Solving this together with the current conservation  \eqref{eq:J=J1=J2}, 
we find a restriction on the shock positions  \cite{bib:BSB} 
 \begin{align}  s_1 +  s_3 = 2 s  ,  \label{eq:s1+s3=s} \end{align} with $s$ \eqref{eq:alpha1=,s=}. 
The form of the density $ \alpha_1  $ is the same as  for the 2-segment case \eqref{eq:alpha1=,s=}.
The current $J$ is also unchanged \cite{bib:BSB}, see fig.~\ref{fig:diagrams}(b). One of the interesting findings in \cite{bib:BSB} is that  we cannot fix  the shock positions $s_1$ and $s_3$ even in the macroscopic level, but they are synchronized  by the restriction \eqref{eq:s1+s3=s}, see fig.~\ref{fig:4-seg} (a).    The right bound of $ s_j $ is 1, and the left  $ \lambda $ is given by solving $   2 s - \lambda  = 1  $: 
\begin{align}   \label{eq:lambda<sj<1}
  \lambda      =   \frac{ 2 ( 1 - 2\rho ) }{ \sqrt{ 1 - r }  }
  <  s_j <  1 .  \end{align} 
The shock positions are uniformly distributed in this range except for the boundaries, see fig.~\ref{fig:4-seg} (b).  Therefore the density profile   by averaging configurations over a large time interval becomes 
\begin{align}\label{eq:rho(x)_SMCSMC_over-time}
\langle  \rho(x) \rangle_{\mathrm T}=
 \begin{cases}  
     \alpha_1  \ (  0  < x < \lambda )   ,    \    \alpha_1    +    (x-\lambda)  g \  (  \lambda  < x < 1   )   , \\ 
        \frac{1}{2} \   (  1 < x < 2  )  ,  \\ 
     \alpha_1  \ (  2  < x < \lambda'   )   ,   \    \alpha_1  +   (x-\lambda')  g \  (  \lambda'  < x < 3    )   , \\ 
        \frac{1}{2} \   (  3 < x < 4   )   ,  
  \end{cases}
\end{align}
where $ \lambda' = \lambda+2 $ and $ g = \frac{ 1 - 2 \alpha_1 }{ 1 - \lambda  } $.  In  fig.~\ref{fig:4-seg} (c), we compare   simulation results of the density profile  $ \rho_i $ \eqref{eq:rho_i:=} with the prediction \eqref{eq:rho(x)_SMCSMC_over-time} with $ x=i /\ell  $. We chose the values of parameters, such that the corresponding densities and left boundaries become 
\begin{align}
  \nonumber 
 &    ( r , \rho   ) = (0.96,  0.5  ) , ( 0.84 , 0.48 ) , ( 0.64,0.44 ) , (0.36,0.38  )   \Leftrightarrow   \\  
 &   ( \alpha_1 , \lambda  ) =  ( 0.4 , 0 ) , (0.3, 0.2   ) ,  ( 0.2 , 0.4  ) ,  ( 0.1, 0.6  ) , 
\end{align}
respectively, according to eqs.~\eqref{eq:alpha1=,s=} and \eqref{eq:lambda<sj<1}. We expect that the discrepancies between the prediction and simulations are  due to the finite-size effect. In the half-filling case $\rho = 1/2$, the profiles of time average in the 1st and 3rd segments are identical to that of the well-studied open TASEP \eqref{eq:rho_i=open}. Therefore one may naively expect that  the 1st and 3rd segments are effectively equivalent to the co-existence line of the open TASEP; e.g. for the 3rd segment, the 2nd and 4th segments play the role of reservoir of densities $\alpha_1$ and $1-\alpha_1$, respectively. Let us observe simulation results of MSD in the 4-segment TASEP, to verify  whether this guess is correct or not.

\begin{figure}\begin{center}  
    \includegraphics[width=0.48\textwidth]{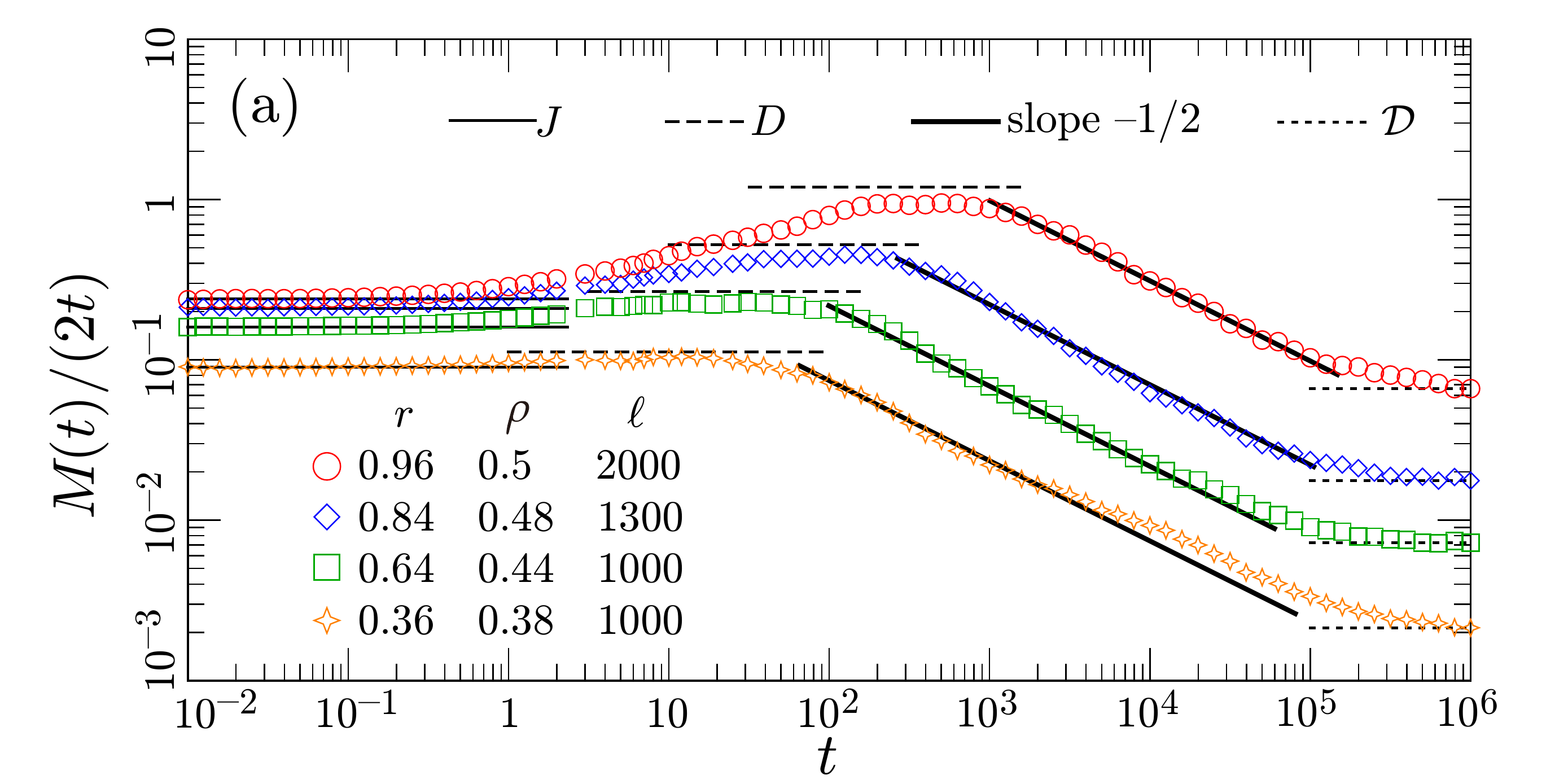}  \ 
    \includegraphics[width=0.24\textwidth]{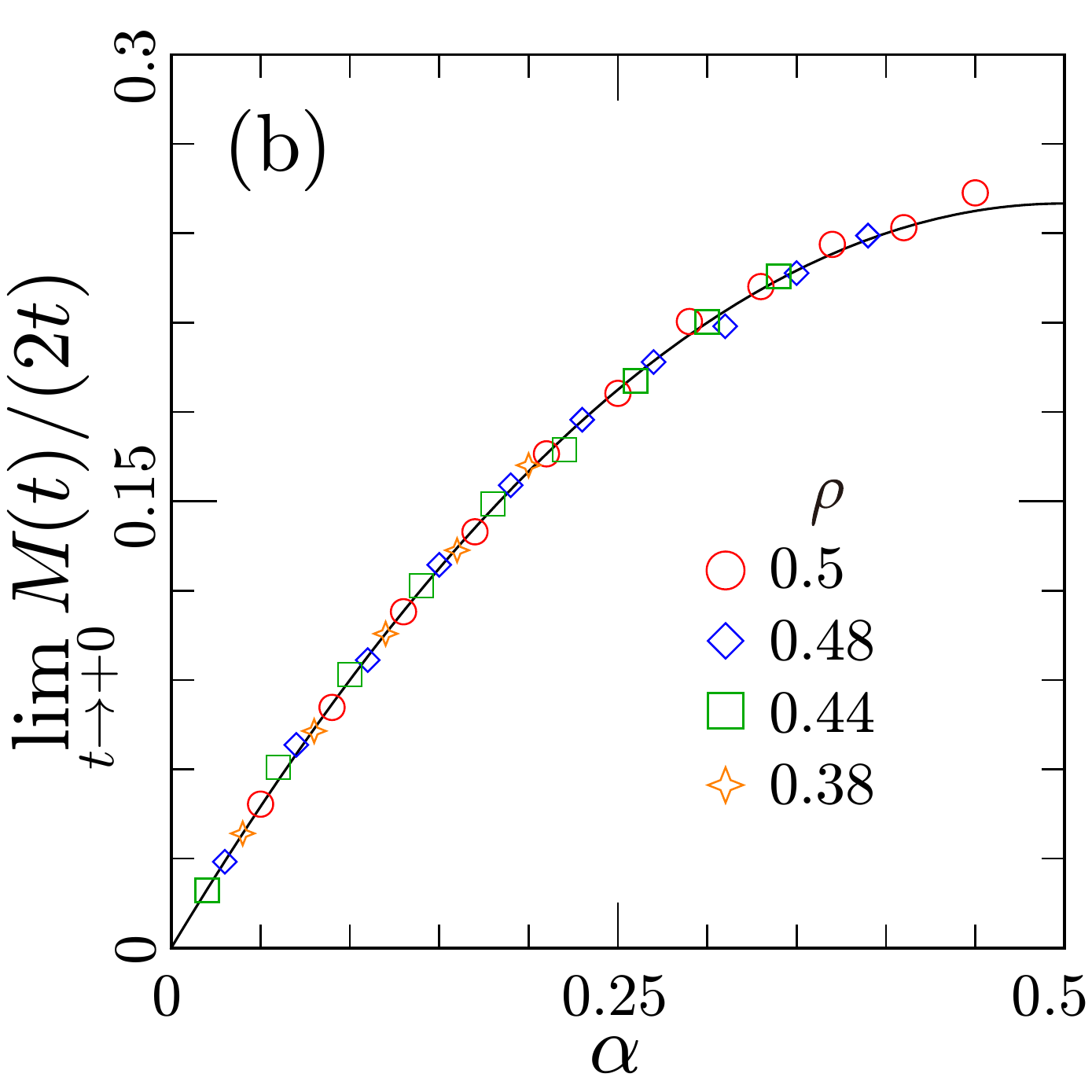}
    \includegraphics[width=0.24\textwidth]{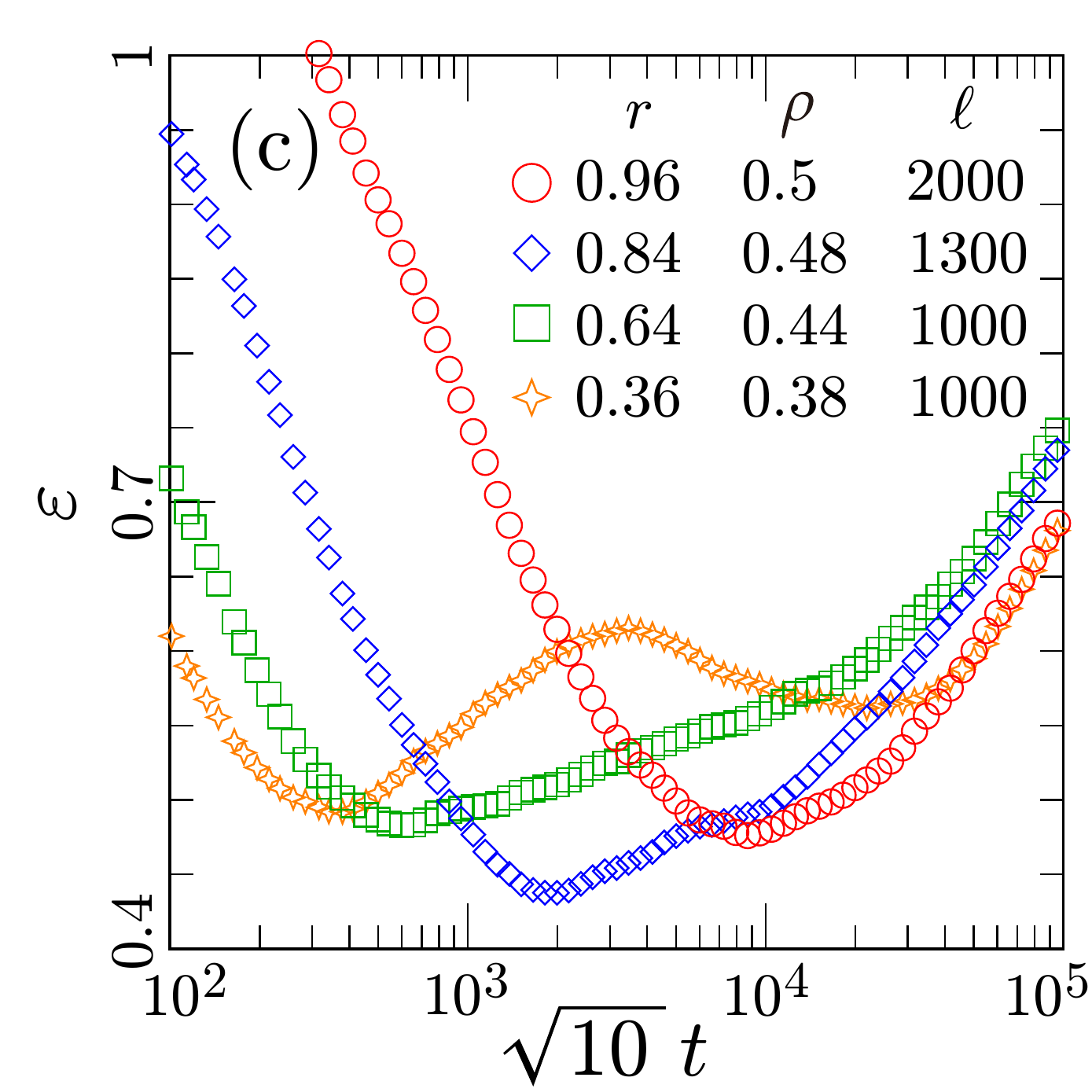}
     \end{center} 
\caption{  
 (a) MSD of the shock positions,
 (b) initial diffusion coefficient, and
 (c) exponent $ \varepsilon$   \eqref{eq:eps=}.        
 In (a), we divided $M(t)$ by $2t$, so as to see its property more easily.
 In (a,b), the solid, dashed and dotted lines correspond to $J$ [eq.~\eqref{eq:J:SMC}], $D$ [eq.~\eqref{eq:D=}] and $\mathcal D$ [simply $M(10^6)  / (2\times 10^6 )  $ from simulations], respectively. 
 To obtain numerical data in (a,c), we took averages over  $ 10^6 $ and $ 10^3 $ simulation runs 
 for $ t\le 2 $ and $ t>2  $, respectively.  For each marker in (b),   we averaged  $  M(t)  / (2t)  $   of   $ 10^{-3} \le t\le  10^{-2}  $  over $ 10^6 $ simulation runs. 
\label{fig:4-MSD} }  
\end{figure}

Denote by $ S_j (t) + (j-1) \ell  $  the microscopic  position  of the second-class particle  in segment  $ j \in \{1,3 \} $ at time $t$, so   $ 1 \le    S_j (t)\le \ell$.   Figure \ref{fig:4-MSD}  (a)  shows the simulation results of the MSDs 
\begin{align} M (t) = \big\langle \big(  S_j (t) - S_j (0) \big)^2 \big\rangle_{\mathrm E} . \end{align}
Thanks to the translational invariance, we omit the subscript $j$, and practically  we obtain $  M(t)  $ by averaging simulation data of $j=1$ and $j=3 $. In some time interval, indeed we see that the diffusion coefficient   \eqref{eq:D=}     with  $\alpha =  \alpha_1 $ \eqref{eq:alpha1=,s=} provides a good estimation of the MSD's behaviour: 
\begin{align} M(t) \approx 2 D ( \alpha_1 ) t . \end{align} 
 Note that the observed  $   M(t) / (2t) $ in this \textit{intermediate diffusive regime} is slightly lower than $ D $, which is expected to be a finite-size effect.

Figure \ref{fig:4-MSD} (a) indicates that the initial and asymptotic behaviours are also diffusive. The current of particles \eqref{eq:J:SMC} gives a good fitting curve for the diffusion coefficient in the \textit{initial diffusive regime}, see fig.~\ref{fig:4-MSD} (b):   
\begin{align} \label{eq:M=2Jt}  M(t) \simeq 2 J  ( \alpha_1  ) \,  t \quad ( t\to +0)  , \end{align} 
which implies that the second-class particles obey the symmetric random walk with jump rate $ J ( \alpha_1  ) $ in a very short time scale.  The same finite-time effect is observed in the open TASEP [fig.~\ref{fig:openTASEP} (a)], where one can prove that  both probabilities of finding configurations $20$ and $12$ are given by $ J(\alpha) $ by using matrices \cite{bib:A1,bib:A2,bib:U,bib:ALS}.

On the other hand,  for the  \textit{asymptotic diffusive regime},
  \begin{align}  M(t) \simeq 2 \mathcal D ( \alpha_1  ) \, t \quad ( t\to \infty) ,    \end{align} 
we expect the inequality   $ \mathcal D ( \alpha_1  ) < D ( \alpha_1  )$,  see fig.~\ref{fig:4-MSD} (a). Note that the conclusion on the asymptotic behaviour is based on our simulations of  up to $t = 10^6$.  Intuitively, no further transition is expected in longer-time simulations. (We considered  the ``asymptotic'' behaviour  in   $ t \ll \ell^2 / \mathcal D $, so that the shocks do not reach  boundaries $ S_j(t) =\lambda \ell , \ell  $. When we take the limit $  t\to \infty $ with $ \ell$ fixed,  the MSD converges to a constant.)  The notation $ \mathcal D (\alpha_1) $ emphasizes that the diffusion coefficient is a function of $\alpha_1$, but the dependence on $ \ell $ and $ \lambda $ may also be nontrivial.

In the  time interval between the intermediate and asymptotic diffusion regimes, the MSD exhibits sub-diffusion.  The bold lines (slope $-1/2 $) are drawn in the log-log graph, fig.~\ref{fig:4-MSD} (a), which agree well with simulations. We also estimate its exponent $\varepsilon$ $(\ln  M(t) \approx  \varepsilon \ln t )$ by 
\begin{align}\label{eq:eps=}
\varepsilon = \frac{ 1 }{ 3t/2 + 1 }
 \sum_{t/2 \le  t' \le 2t } \frac{ \ln M ( 10t' ) - \ln M ( t' ) }{ \ln 10 } , 
\end{align}
in this \textit{transient regime.}  
The exponent by this formula takes the minimum, which is  $\approx 1/2$, see  fig.~\ref{fig:4-MSD} (c).

\begin{figure}\begin{center} 
     \includegraphics[width=0.24\textwidth]{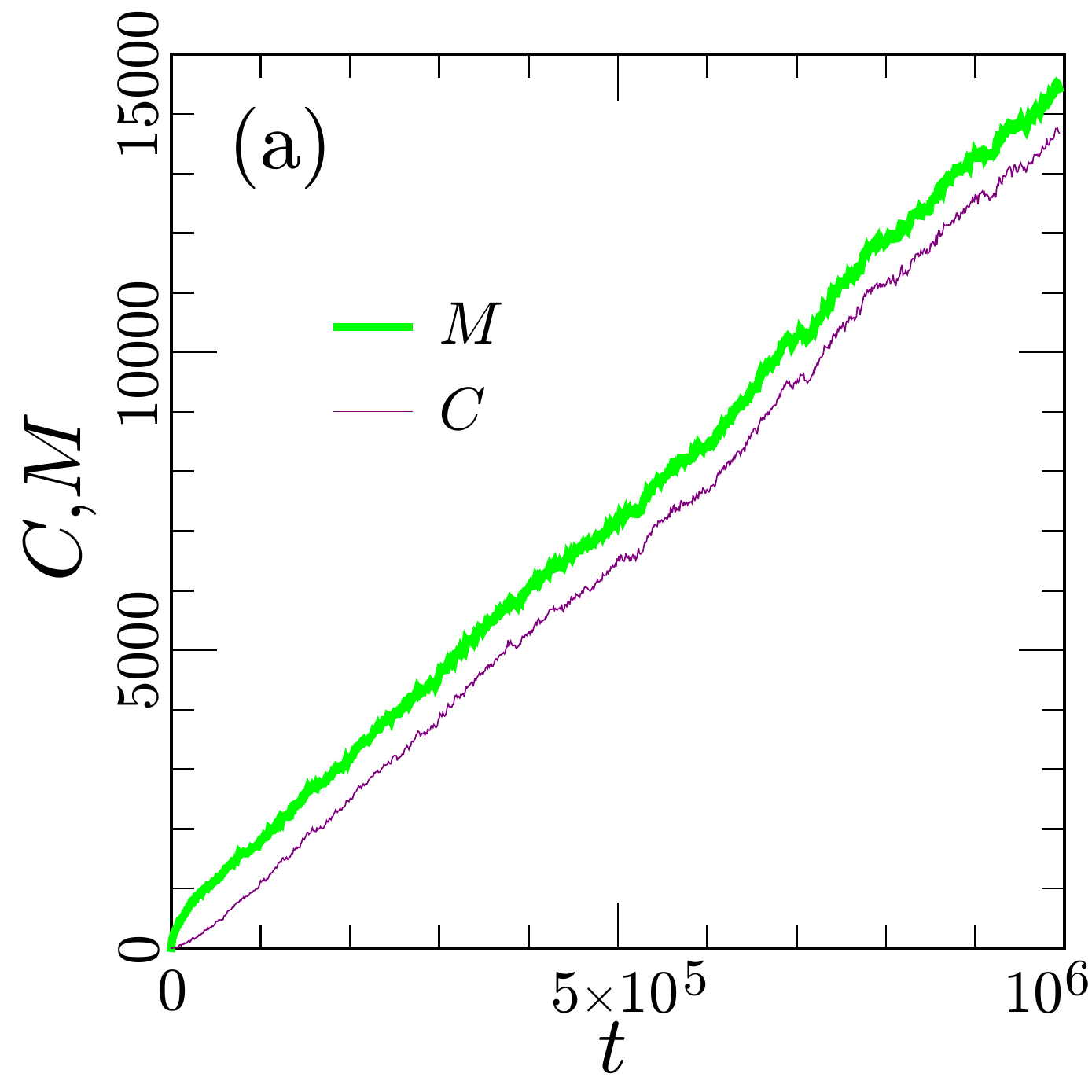} \ 
     \includegraphics[width=0.24\textwidth]{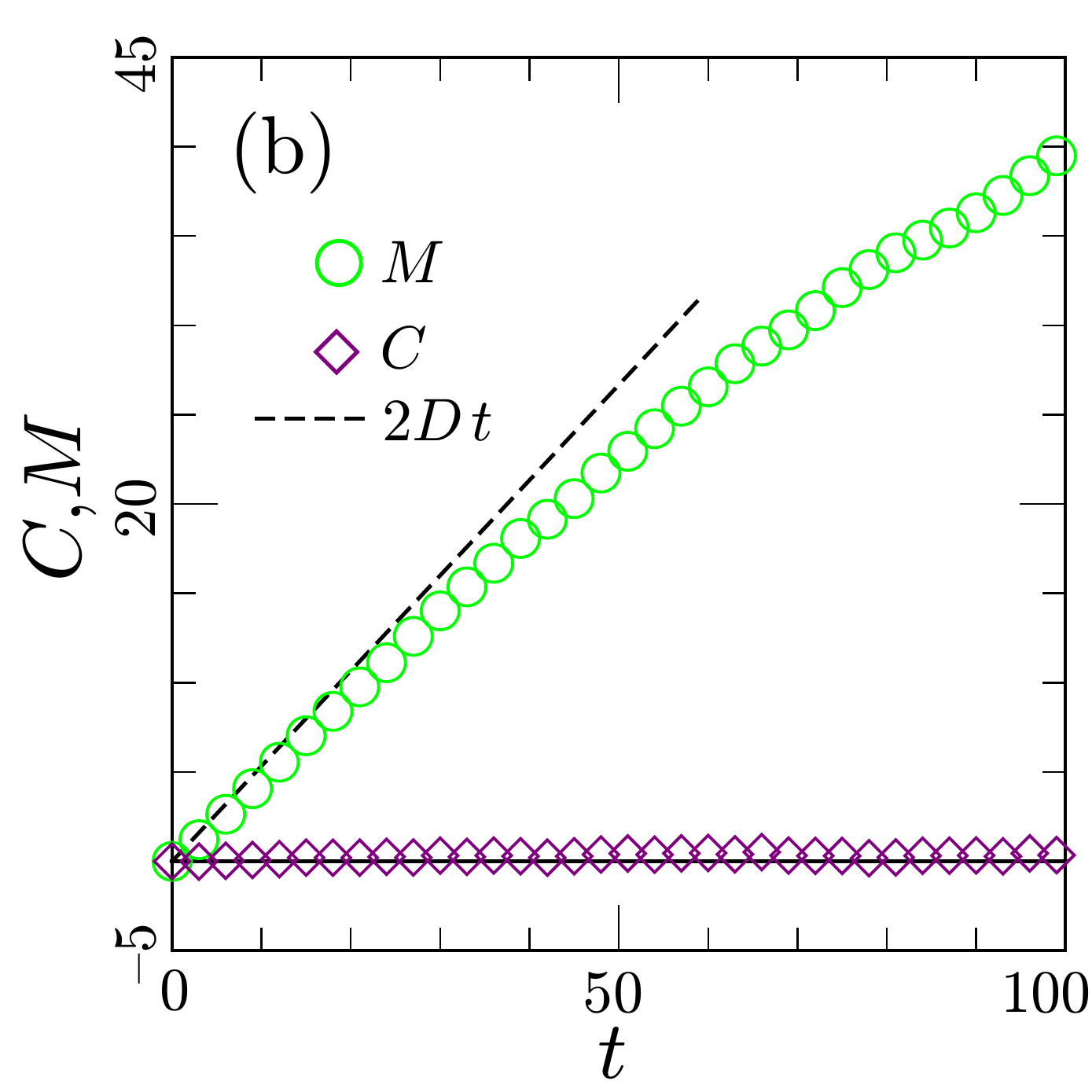} \ 
     \includegraphics[width=0.24\textwidth]{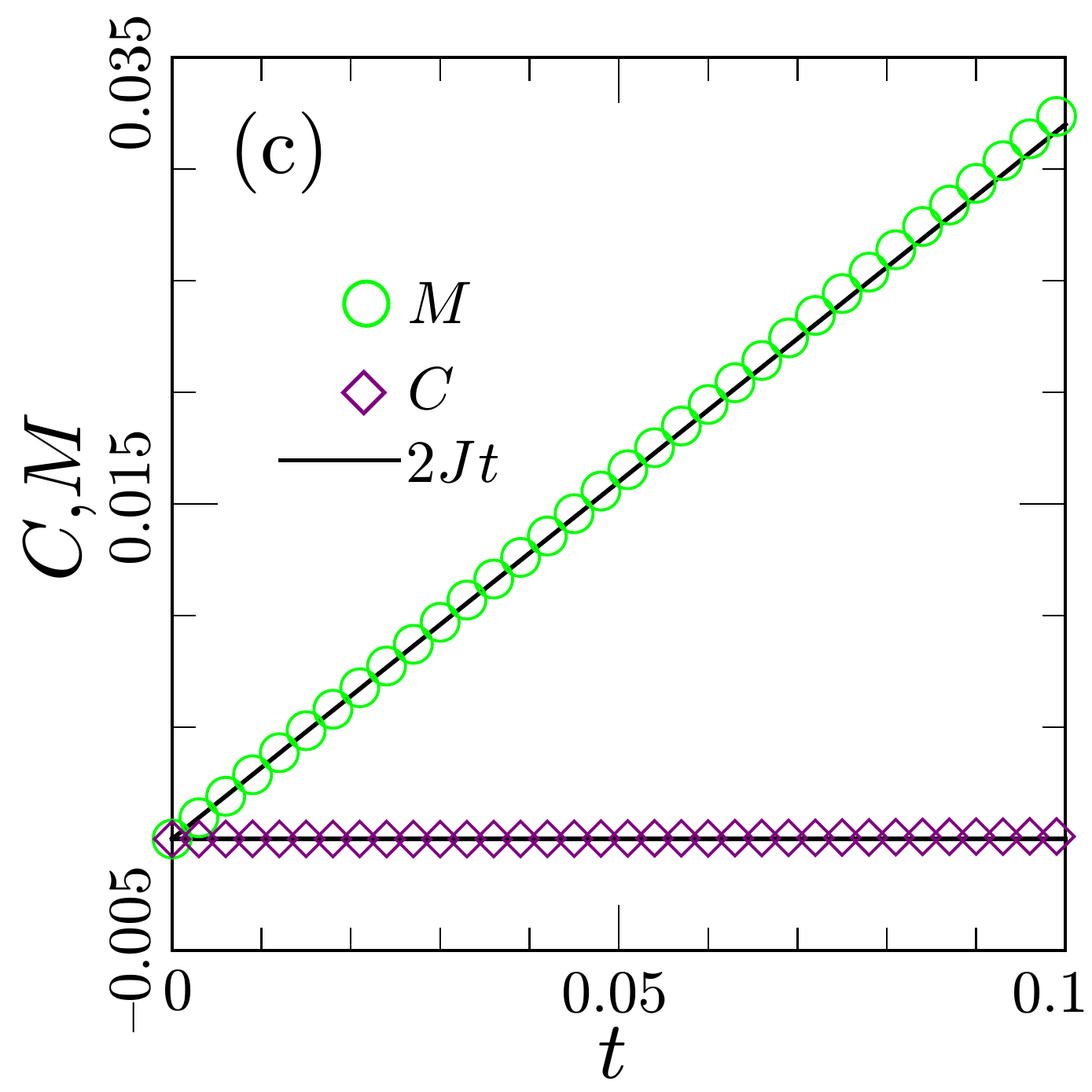} \ 
     \includegraphics[width=0.24\textwidth]{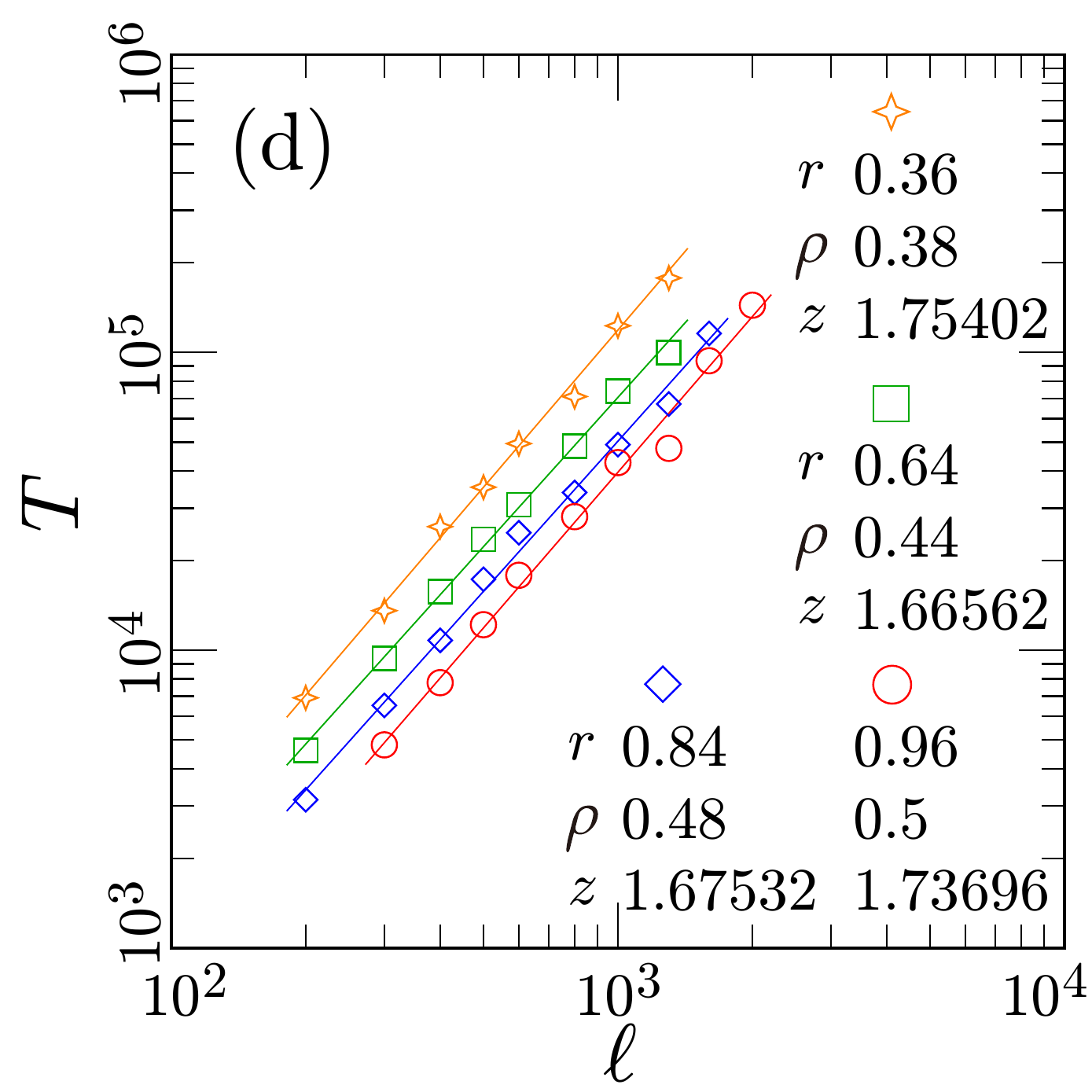}  \end{center} 
 \caption{  
 (a,b,c) Comparison between the MSD and the correlation function  
  in different time windows, and
  (d) crossover time vs segment length. 
   For  (a,b,c),  we have set the values of the parameters as $ (\ell,r, \rho) = (1000,0.64,0.44) $,  
   and averaged over $ 10^3,10^4 $ and $ 10^6 $ simulation runs,  respectively.
 In (d), each marker was plotted by averaging over  $ 10^3 $ simulation runs, 
  and the lines are fitting curves $ c \, \ell^z$.  
\label{fig:4-Correlation} } 
\end{figure}

Let us investigate the correlation of the shocks.  We expect that, in a very short time,  the two shocks move independently, since they are far from each other. On the other hand,   they are  synchronized in the frame of  a larger time scale, e.g. as  fig.~\ref{fig:4-seg} (a). In order to  quantify  (in)dependency of the two shocks, we introduce a correlation function 
\begin{align}
\label{eq:C(t)=}
 C(t)  = - \big\langle   \big( S_1 (t) - S_1  (0) \big) \big( S_3  (t) - S_3  (0)   \big)  \big\rangle_{\mathrm E}   . 
\end{align}
Due the synchronization $ S_1 (t) + S_3 (t) \approx \ell s $, we have  
\begin{align} C(t) \simeq M (t) \quad ( t \to + \infty), \end{align}  
   which is observed in fig.~\ref{fig:4-Correlation} (a).
On the other hand, their initial behaviours are completely different, see fig.~\ref{fig:4-Correlation} (b,c).  In the vicinity of $t= 0$, we conjecture that $C(t)$ increases more slowly than any power function $t^a (a>0)$.

Let us define the time $T$ as 
 \begin{align} T  = \inf \big\{ t > 0 \big|  2  C(t)   > M(t)  \big\} ,  \end{align}  
characterizing the crossover between the time scales of independency and synchronization.  
Figure \ref{fig:4-Correlation} (d) shows  $T$ vs segment length $\ell$. By assuming the power law  $ T \simeq c\, \ell^z $, we perform fitting from the simulation data that we have. We write the results directly in fig.~\ref{fig:4-Correlation} (d), and draw corresponding straight lines as well. 
 Under the further assumption that the exponent is independent from parameters, 
we simply average the four obtained exponents:  $z \approx 1.71$.

\section{Discussions}\label{sect:disc}
In this work,  we investigated synchronization of shocks in the 4-segment  TASEP, by means of  the two second-class particles.  We found that the behaviour of the MSD $ M(t) $ of the shocks is not so simple. In the initial regime (very short time scale), it is diffusive and the coefficient is nothing but the formula for the particle current. In the intermediate diffusive regime, the coefficient $D$ known in the open TASEP provides a good estimation. This fact indicates that,  up to the intermediate diffusion regime,  the shocks' motions are determined by the values of low and high densities, and independent from details of boundary conditions.  Via sub-diffusive regime, the MSD achieves the asymptotic diffusive regime, where the diffusion coefficient is different from $D$.  From the log-log graph and the numerical estimation, fig.~\ref{fig:4-MSD} (a,c), it seems that $ M(t)\propto \sqrt t$ in the sub-diffusive regime.   In some random walks,   similar changes of the diffusivity have been found \cite{bib:PM,bib:TSJS}.    In our case, the \textit{regime change} is spontaneously induced by the particle number conservation, without directly imposing any interaction between the two second-class particles in defining the model.

We also investigated the correlation function of the shocks, and the crossover time between independency and synchronization of the shocks. The correlation function $C(t)$ \eqref{eq:C(t)=} becomes identical to $M(t)$ as $t\to +\infty$.  In the vicinity of $  t=0 $, $ C(t) $ increases very slowly, whereas  $ M(t) $ is linear.  We defined the crossover time $T$ as the time when the ratio $ C/M $ exceeds $ 1/2 $.   We estimated the dynamical exponent $z$ for $ T $ by using our simulation data, which was found between  the KPZ $ z=3/2  $ \cite{bib:GS} and the normal diffusion $ z=2 $.  The  exponent $z$ as well as $ \mathcal D  $ and $ \epsilon $ should be  more precisely estimated  in larger systems   with longer simulation time and by more simulation runs.

The generalization to the $2n $-segment case is straightforward: the $j$th segment has the rate 1 (resp. $r$) when $j$ is odd (resp. even). When $ \rho < \rho_c $, the density profile $ \rho (x)  \ ( j - 1 < x < j ) $ of $j$th segment is give as $\rho (x)  = \alpha_1$ for odd $j$, and $\rho (x)  = \alpha_2$ for even $j$, with the same forms as in the 2-segment case \eqref{eq:alpha1=,alpha2=}. When the global density $ \rho_c < \rho < 1- \rho_c $, the density profiles are flat with density $ 1/2 $ in the segments with even numbers.  On the other hand, $n$ shocks appear in segments $j=1,3, \dots,2n-1 $. Denoting their positions by $ s_j + j - 1 $, we find 
 $ s_1 + s_3 + \cdots + s_{2n-1 } = n s ,  $ from the conservation of the number of particles. As an analogue to the $n=2$ case, we expect that they are not static but synchronized. Since only one equation governs the synchronization of the $ n $ shocks, we naturally expect that the asymptotic diffusion constant $ \mathcal D_n( \alpha )$ enjoys 
 $   \mathcal D_1 (  \alpha ) < \mathcal  D_2 (  \alpha ) < \mathcal  D_3 (  \alpha ) <  \cdots $
 with $ \mathcal D_1 \equiv 0 $ and $ \mathcal D_2 ( \alpha ) = \mathcal D ( \alpha ) $. 
A further intuitive conjecture is that the sub-diffusive regime vanishes as $ n \to + \infty $, and 
$\lim_{ n\to\infty } \mathcal  D_n (  \alpha )   = D(\alpha)$. 
 
 It is known that two shocks are synchronized in the model \cite{bib:CCB}. A simple generalization of the JL model \cite{bib:SB}, e.g. $ p_i = r ( i \in \{\ell , 2\ell\} ) , 1 ( i \notin \{\ell , 2\ell\} ) $ with $L =2 \ell $, also exhibits the same type of synchronization.  One of important questions is whether the MSDs of shocks in these models behave like fig.~\ref{fig:4-MSD} (a), and if yes,  whether  the asymptotic  diffusion coefficient is the same as for the 4-segment TASEP.   Note that the positions of the two synchronized shocks are, in general, far from each other in these models as well as the 4-segment TASEP.  (See e.g.  \cite{bib:MH,bib:JWW,bib:J,bib:DG} for other types of  synchronization.) While the motions of the  second-class particles  are locally defined, the shocks  move as if they \textit{knew} each other's position.    We believe that this viewpoint gives hints to study self-organization phenomena,  e.g. in biological cells,  as the exclusion process is one of basic models also in biophysics \cite{bib:CMZ}.  Application to traffic flows would be also an interesting problem. The local inhomogeneities in the JL and generalized JL models are very similar to traffic lights, and combinations of segments that we studied here evoke different limit speeds of cars.

\section*{Acknowledgements}
 The author thanks Ludger Santen and M Reza Shaebani for useful discussions.

\end{document}